\pdfoutput=1
\documentclass[twocolumn,showpacs,preprintnumbers,superscriptaddress,prb,amsmath,amssymb]{revtex4-1}

\usepackage{graphicx}% Include figure files
\usepackage{dcolumn}% Align table columns on decimal point
\usepackage{bm}% bold math
\usepackage{xcolor}
\usepackage{ulem}
\usepackage{gensymb}
\usepackage{braket}
\usepackage{epstopdf}
\usepackage{amsmath}
\usepackage{xcolor}
\usepackage{mathtools}
\usepackage[percent]{overpic}
\usepackage{enumitem}
\usepackage{mhchem}

\begin{document}

\definecolor{mycyan}{cmyk}{1,0,0,0.3}
\definecolor{mymagenta}{cmyk}{0,1,0,0.3}
\definecolor{myyellow}{cmyk}{0,0,1,0.4}
\definecolor{mygray}{cmyk}{0,0,0,0.8}

\newcommand{\cre}[2]{{#1}_{#2}^{\dagger}}
\newcommand{\ann}[2]{{#1}_{#2}^{\vphantom{\dagger}}}
\newcommand{\veck}[1]{\boldsymbol{#1}}

\newcommand{\dif}{\mathrm{d}}
\newcommand{\pard}[1]{\frac{\partial}{\partial #1}}
\newcommand{\pcre}[1]{\cre{\phi}{#1}}
\newcommand{\pann}[1]{\ann{\phi}{#1}}

\newcommand{\figref}[1]{Fig.~\ref{#1}}

\title{Orbital and spin order in spin-orbit coupled $d^1$ and $d^2$ double perovskites}

\author{Christopher Svoboda}
\author{Mohit Randeria}
\author{Nandini Trivedi}
\affiliation{Department of Physics, The Ohio State University, Columbus, Ohio 43210, USA}
\date{\today}

\begin{abstract}
We consider strongly spin-orbit coupled double perovskites A$_2$BB'O$_6$ with B' magnetic ions in either $d^1$ or $d^2$ electronic configuration and non-magnetic B ions. We provide insights into several experimental puzzles, such as the predominance of ferromagnetism in $d^1$ versus antiferromagnetism in $d^2$ systems, the appearance of negative Curie-Weiss temperatures for ferromagnetic materials and the size of effective magnetic moments.
We develop and solve a microscopic model with both spin and orbital degrees of freedom within the Mott insulating regime at finite temperature using mean field theory.
The interplay between anisotropic orbital degrees of freedom and spin-orbit coupling results in complex ground states in both $d^1$ and $d^2$ systems.
We show that the ordering of orbital degrees of freedom in $d^1$ systems results in coplanar canted ferromagnetic and 4-sublattice antiferromagnetic structures. In $d^2$ systems we find additional colinear antiferromagnetic and ferromagnetic phases not appearing in $d^1$ systems.
At finite temperatures, we find that orbital ordering driven by both superexchange and Coulomb interactions may occur at much higher temperatures compared to magnetic order and leads to distinct deviations from Curie-Weiss law. 
\end{abstract}

%\pacs{
%75.10.Jm, Quantized spin models, including quantum spin frustration
%71.70.Ej  Spin-orbit coupling, Zeeman and Stark splitting, Jahn-Teller effect
%75.30.Et  Exchange and superexchange interactions
%71.27.+a  Strongly correlated electron systems; heavy fermions
%75.47.Lx  Magnetic oxides
%}

\maketitle

\section{Introduction}

%The parent compounds of the high $T_c$ copper-based superconductors can effectively be described by one hole in the $e_g$ orbitals. Given the Jahn-Teller distortion of these doubly degenerate orbitals we then obtain a single orbital Hubbard model with one electron per orbital as the model Hamiltonian. As is well known, for large on-site interactions this Hubbard model reduces to a $S=1/2$ quantum antiferromagnetic Heisenberg model on a square lattice. 
%In this paper we consider the analogous situation in the triply degenerate $t_{2g}$ orbitals. XXXwhy eg cannot have SOCXXX
%Given the effective $L=1$ algebra of $t_{2g}$ orbitals, this allows us to investigate the role of spin-orbit coupling (SOC) and Coulomb correlations.
%***********
Strong  SOC in correlated materials has provided a platform for quantum spin liquids, Weyl semimetals, and an ongoing search for high $T_c$ superconductivity in the iridates.
\cite{KrempaBalents2014,RauLeeKee2016}
Among the strongly spin-orbit coupled materials include $4d$ and $5d$ double perovskites A$_2$BB'O$_6$ with electron counts $d^1$-$d^5$ on the magnetic B' ion. Here we restrict our discussion to $B$ site ions that are non-magnetic.
Due to large distances between the magnetic ions, these materials are often Mott insulators and present a promising class of materials to explore the interplay of strong correlations and spin-orbit coupling.
Additionally, the magnetic sites form an FCC lattice leading to frustrated magnetism.

Each electron count carries a different total angular momentum quantum number providing a new platform for studying novel magnetism.
The half filled $t_{2g}$ shells of $d^3$ ions result in an effective spin-3/2 model which are nominally expected to be described as a classically frustrated spin systems.
\cite{KermarrecGaulin2015,TaylorChristianson2016}
In the opposite limit, $d^5$ systems with $j=1/2$ are both intrinsically more quantum and are protected from local distortions by time reversal symmetry.
These systems may offer a route to realizing Kitaev physics and more generally spin liquids in three dimensions.
\cite{CookParamekantiFCC2015,CookParamekantiFCC2016}
The $d^4$ case is especially unique since spin-orbit coupling dictates that local moments should be absent and magnetism is forbidden.
However several theoretical \cite{d4Khaliullin2013,Akbari2014,AritaKunes2016,Meetei2015} and experimental \cite{Buchner2016,Kennedy2015,PhelanCava2016} studies have examined the possibility of inducing local moments through superexchange interactions.

The $d^1$ and $d^2$ electron counts stand out in that they combine aspects of the former three electron counts and will be the focus of this paper.
First, they possess local angular momenta large enough to support quadrupolar order.
Second, they possess unquenched orbital degrees of freedom that result in highly anisotropic interactions between magnetic ions.\cite{KugelKhomskii,KhaliullinPTPS}
Both of these aspects will allow for the orbital degrees of freedom to play a significant role in determining the spin, orbital, and spin-orbital ordering.

% \subsection{Experiments on $d^1$ and $d^2$ systems}
While both electron counts have similar potential, experimental observations of magnetic properties of the $d^1$ double perovskites have drawn significant interest.
The $4d^1$ compound \ce{Ba2YMoO6} shows no long range magnetic order down to 2K despite having a large Curie-Weiss temperature $\theta = -160K$ and retaining cubic symmetry which leads to the conclusion that the ground state consists of valence bonds \cite{deVries2010,Aharen2010d1,Carlo2011,deVries2013,QuZhang2013}.
Among the $5d^1$ compounds are ferromagnetic \ce{Ba2NaOsO6} \cite{Stitzer2002,EricksonFisher2007,Steele2011,Mitrovic2017}, \ce{Ba2MgReO6} \cite{Bramnik2003,Marjerrison2016}, and \ce{Ba2ZnReO6} \cite{Marjerrison2016} which is unusual since ferromagnetism in Mott insulators is uncommon.
There are two additional twists to the story: first, negative Curie-Weiss temperatures have been observed in these ferromagnets, and, second, \ce{Ba2LiOsO6} is antiferromagnetic despite sharing the same cubic structure as \ce{Ba2NaOsO6}.\cite{Stitzer2002}
The $d^2$ compounds offer a similar platform to search for unusual magnetism, however experimental studies seem to suggest that antiferromagnetic interactions are more prevalent in $d^2$ systems.
Phase transitions to antiferromagnetic order are reported in \ce{Ca3OsO6} \cite{Feng2013}, \ce{Ba2CaOsO6} \cite{Thompson2014}, and \ce{Sr2MgOsO6} \cite{Yuan2015,Morrow2016} while glass-like transitions are reported in \ce{Ba2YReO6} \cite{Aharen2010d2}, \ce{Ca2MgOsO6} \cite{Yuan2015}, and \ce{Sr2YReO6} \cite{Aczel2016}.
There are also several alleged singlet ground states: \ce{La2LiReO6} \cite{Aharen2010d2}, \ce{SrLaMReO6} \cite{Thompson2015}, and \ce{Sr2InReO6} \cite{Aczel2016}.

% \subsection{Previous theoretical work}
Many theoretical investigations have been undertaken to understand the magnetism in both $d^1$ and $d^2$ double perovskites.
In the limit of large spin-orbit coupling, the spin $S = 1/2$ and orbital $L_\mathrm{eff} = -1$ angular momenta add to a total angular momenta of $j = 3/2$.
Within the $jj$-coupling scheme, magnetic moments are identically zero due to cancellation of the spin and orbital moments, $\veck{M} = 2\veck{S} - \veck{L} = 0$.
On the other hand, $d^2$ systems allow for a nonzero moment of $\veck{M} = \tfrac{\sqrt{6}}{2} \mu_B \veck{J}$ for total $J=2$ within the $LS$-coupling scheme.
However both systems are experimentally observed to be magnetic.
Density functional theory studies have recently revealed the importance of oxygen hybridization in suppressing the orbital moment so that a large non-zero moment results.\cite{XuBrink2016,AhnKunes2016}
Other density functional theory studies have pointed out that spin-orbit coupling and hybridized orbitals play a major role in opening a gap within DFT+U.\cite{Xiang2007,LeePickett2007,GangopadhyayPickett2015}

Model approaches have shed some light on the nature of the magnetically ordered states by using spin-orbital Hamiltonians \cite{RomhanyiBalentsJackeli2016}, projecting spin-orbital Hamiltonians to the lowest energy total angular momentum multiplet \cite{ChenBalents2010,ChenBalents2011} or lowest energy doublet \cite{DoddsYBKim2011}, and other approaches \cite{IshizukaBalents2014}.
In both electron counts, Chen et. al.\cite{ChenBalents2010,ChenBalents2011} find canted ferromagnetism accompanied by quadrupolar order occupies a majority of parameter space.
Additionally they find a novel non-colinear antiferromagnetic phase in $d^2$, but not $d^1$, which was recently found in $d^1$ as the most energetically favorable antiferromagnetic state.\cite{RomhanyiBalentsJackeli2016}
Proposals for both valence bond ground states\cite{ChenBalents2010,RomhanyiBalentsJackeli2016} and quantum spin liquids\cite{ChenBalents2010,NatoriPereira2016} also exist.

Yet many puzzles remain unsolved.
Despite predictions for canted ferromagnetic phases\cite{ChenBalents2010,ChenBalents2011} in both $d^1$ and $d^2$, many ferromagnetic $d^1$ systems exist but few $d^2$ ferromagnets exist.
Furthermore, the physical origin of negative Curie-Weiss temperatures in these ferromagnets is still not understood, and there are multiple studies trying to reproduce the magnitudes of the effective Curie moments experimentally measured.

% \subsection{Our main results}
Here we study magnetic models for the $d^1$ and $d^2$ cubic double perovskites with strong spin-orbit coupling with both spin and orbital degrees of freedom at finite temperature.
While we are focusing on applications to ordering in $5d$ cubic double perovskites, our results may also apply to $4d$ compounds and non-cubic double perovskites as well.
Despite the greater complexity than the $J=3/2$ and $J=2$ multipolar descriptions, the spin-orbital picture actually leads to an intuitive and qualitative understanding of several aspects of the phenomenology in these double perovskites.
In our study of magnetically ordered phases, we arrive at several conclusion which we now list.

First, our results emphasize the importance of the orbital degrees of freedom and anisotropic interactions that accompany them.
In particular, we show that the anisotropic interactions result in orbital order that stabilizes exotic magnetic order.
The orbital (quadrupolar) ordering temperature scale is set both by superexchange interactions and by inter-site Coulomb repulsion, and, in several cases, the orbital ordering temperature can be much larger than the magnetic ordering temperature.

Second, although we start with the same electronic model for both $d^1$ and $d^2$ systems, the energetics of the ground states strongly depend on the electron count.
This is reflected in how the spin and orbital degrees of freedom order and provides a qualitative understanding for why ferromagnetism has been repeatedly observed in $d^1$ systems while antiferromagnetic interactions remain prevalent in $d^2$ systems.

Third, the onset of orbital order causes changes in magnetic susceptibility resulting in non-Curie-Weiss behavior.
Our model gives the appearance of a negative Curie-Weiss temperature for the ferromagnetic phase while still retaining a properly diverging susceptibility at the ferromagnetic transition.

Fourth, if orbital order occurs, hybridization with oxygen alone does not reproduce the experimentally determined values of the magnetic moments in $d^1$ systems.
Corrections are necessary which may arise from dynamical Jahn-Teller effects\cite{XuBrink2016} and more generally with mixing of the $j=3/2$ and $j=1/2$ states as we propose.
Charge transfer from oxygen might also be considered for systems where the Curie moment has measured be in excess of 1$\mu_B$.\footnote{Private communications.}

Lastly, we outline where our calculations stand with respect to other work.
First, our zero temperature phase diagram for $d^1$ contains a 4-sublattice antiferromagnetic phase and a canted ferromagnetic phase which share underlying orbital ordering patterns.
Our findings are compatible with those of Romh\'{a}nyi et. al. \cite{RomhanyiBalentsJackeli2016}, and we further provide a clear interpretation of why these orbital ordering patterns occur, how they dictate the magnetic ordering, and then extend our calculations to finite temperature.
Like Chen et. al., we find that orbital ordering can occur at temperatures much higher than the magnetic ordering temperature, however, it leads to a different interpretation of the negative Curie-Weiss temperature in $d^1$ ferromagnets.
Furthermore, our spin-orbital approach includes mixing between the $j=3/2$ and $j=1/2$ states induced by orbital order and intermediate spin-orbit coupling energy scales.
Second, our zero temperature phase diagram for $d^2$ differs remarkably from that of Chen et. al.\cite{ChenBalents2011} which we discuss in detail in later sections.
However, the most significant difference is in the energetics of antiferromagnetism versus ferromagnetism which gives a qualitative explanation for the broadly observed differences in ordering between $5d^1$ and $5d^2$ compounds.
Finally, we do not consider valence bond or spin liquid phases in this work although both may be applicable to $d^1$ and $d^2$ systems.

Experimentally, many of our findings can be tested using multiple probes.
At the orbital ordering temperature, there will be second order phase transition with a signature in heat capacity as well as changes in the magnetic susceptibility which are relevant for both powder samples and single crystals.
NMR/NQR has recently found evidence of time-reversal invariant order above the magnetic ordering temperature in \ce{Ba2NaOsO6}.\cite{Mitrovic2017} 
Resonant X-ray scattering may also provide crucial insights into this hidden order as it is sensitive to orbital occupancy.
We show that time-reversal invariant orbital order occurs in both ferromagnetic and antiferromagnetic phases we find, and we suggest that experimental probes which are sensitive to such order should also be pointed at the antiferromagnetic compounds as well.

\section{$d^1$ Double Perovskites}

Here we develop a spin-orbital model for the $d^1$ double perovskites with magnetic B' ions with spin-orbit coupling featuring both spin-orbital superexchange and inter-site Coulomb repulsion between B' ions.
We then solve the model within mean field theory at both zero temperature and finite temperature.
At zero temperature, we find phases with orbital order and show how this ordering restricts the magnetic order.
At finite temperature, we examine how orbital order modifies magnetic susceptibility and the Curie-Weiss parameters.

\subsection{Model}
In the presence of cubic symmetry, the magnetic $B'$ ions form an FCC lattice and contain one electron in the outermost $d$ shell.
The five degenerate levels are split by the octahedral crystal field into the higher energy $e_g$ orbitals and lower $t_{2g}$ orbitals so that the $t_{2g}$ shell contains one electron.
The electronic structure for the $t_{2g}$ orbitals may be approximated by a nearest neighbor tight-binding model where only one of the three orbitals interacts along each direction.
\begin{equation}
H_\mathrm{TB} = -t \sum_{\alpha} \sum_{\langle ij \rangle \in \alpha} \cre{c}{i,\alpha} \ann{c}{j,\alpha} + \mathrm{h.c.}
\end{equation}
Here the sum over $\alpha$ is over all $yz$, $zx$, and $xy$ planes in the FCC lattice.
For B' sites in plane $\alpha$, the $\alpha$ orbital on site $i$ overlaps with the $\alpha$ orbital on site $j$.
Each $\alpha$ orbital has four neighboring $\alpha$ orbitals in its plane plane giving a total of twelve relevant B' neighbors per B' site.
In addition to the tight-binding term, the unquenched $t_{2g}$ orbital angular momentum $L=1$ results in a spin-orbit coupling on each $B'$ ion\cite{RauLeeKee2016} $H_\mathrm{SO} = -\lambda \sum_i \veck{L}_i \cdot \veck{S}_i$.
Here the orbital $L=1$ and spin $S=1/2$ operators both satisfy the usual commutation relations for angular momentum (ie. $\veck{L} \times \veck{L} = i \hbar \veck{L}$).

The on-site multi-orbital Coulomb interaction is given by $H_\mathrm{U} = \sum_i H_\mathrm{U}^{(i)}$
\begin{equation}
H_\mathrm{U}^{(i)} = \tfrac{U-3J_H}{2}  N_i (N_i - 1) + J_H \left( \tfrac{5}{2} N_i - 2\veck{S}_i^2 - \tfrac{1}{2} \veck{L}_i^2 \right)
\end{equation}
where $U$ is the Coulomb repulsion and $J_H$ is Hund's coupling. \cite{GeorgesMediciMravlje2013}
Being in the Mott limit, we calculate the effective spin-orbital superexchange Hamiltonian within second order perturbation theory.
The superexchange Hamiltonian is given by the following
\begin{equation}
\begin{split}
H_\mathrm{SE} = -\frac{J_\mathrm{SE}}{4} \sum_\alpha \sum_{\langle ij \rangle \in \alpha} \left\lbrace r_1 (\tfrac{3}{4} + \veck{S}_i \cdot \veck{S}_j) (n_i^\alpha - n_j^\alpha)^2  \right. 
\\ \left. + (\tfrac{1}{4} - \veck{S}_i \cdot \veck{S}_j) \left[ r_2 (n_i^\alpha + n_j^\alpha)^2 + \tfrac{4}{3} (r_3 - r_2) n_i^\alpha n_j^\alpha  \right]  \right\rbrace 
\end{split}
\label{HSEd1}
\end{equation}
where $J_\mathrm{SE} = 4 t^2 / U$ is the superexchange strength and $r_1 = (1-3\eta)^{-1}$, $r_2 = (1-\eta)^{-1}$, and $r_3 = (1+2\eta)^{-1}$ with $\eta = J_H / U$.\cite{KhaliullinPTPS}
Here the $t_{2g}$ orbital electron occupation numbers are written as $n_i^\alpha = \cre{c}{i,\alpha} \ann{c}{i,\alpha}$
The top line of equation \eqref{HSEd1} contributes a ferromagnetic spin interaction which requires that one of the two orbitals along a bond is occupied while the other is unoccupied.
The bottom line of equation \eqref{HSEd1} contributes an antiferromagnetic spin interaction which is maximized when both orbitals along a bond are occupied.
The strength of Hund's coupling, $J_H / U$, determines the strength of the two interactions relative to each other.

Due to the large spatial extent of $5d$ orbitals from strong oxygen hybridization, we include a term accounting for the Coulomb repulsion between orbitals on different
sites.\cite{ChenBalents2010}
Let $(\alpha,\beta,\gamma)$ be a cyclic permutation of the $t_{2g}$ orbitals $(yz,zx,xy)$.
The repulsion is described by the following:
\begin{equation}
\label{HVeqn}
H_\mathrm{V} = V \sum_\alpha \sum_{\langle ij \rangle \in \alpha} \left[ \tfrac{9}{4} n_i^\alpha n_j^\alpha - \tfrac{4}{3} (n_i^\beta - n_i^\gamma)(n_j^\beta - n_j^\gamma) \right]
\end{equation}
While the coefficients in \eqref{HVeqn} are only quantitatively correct in the limit of quadrupolar interactions, they qualitatively capture the correct repulsion.
For example, within the $xy$ plane, a pair of $xy$ orbitals repel each other more than an $xy$ and $yz$ orbital.

The total effective magnetic interaction then reads $H=H_\mathrm{SO} + H_\mathrm{SE} + H_\mathrm{V}$.
Of the three parameters, spin-orbit coupling has the largest energy scale $\lambda \approx 0.4\,\mathrm{eV}$ for the $5d$ oxides while superexchange and and intersite Coulomb repulsion are taken to have energy scales on the order of tens of meV.
For $4d$ oxides, the spin-orbit energy scale is reduced to $0.1-0.2\,\mathrm{eV}$ so that mixing between the $j=3/2$ and $j=1/2$ states is likely to occur.
While our spin-orbit superexchange interaction is calculated in the $LS$-coupling scheme, recent evidence suggests that the true picture for the $5d$ oxides lies between the $LS$ and $jj$ limits.\cite{YuanYJKim2017}

\begin{figure*}
\includegraphics[width=\textwidth,trim={0.8cm 0.5cm 0.8cm 0.8cm}]{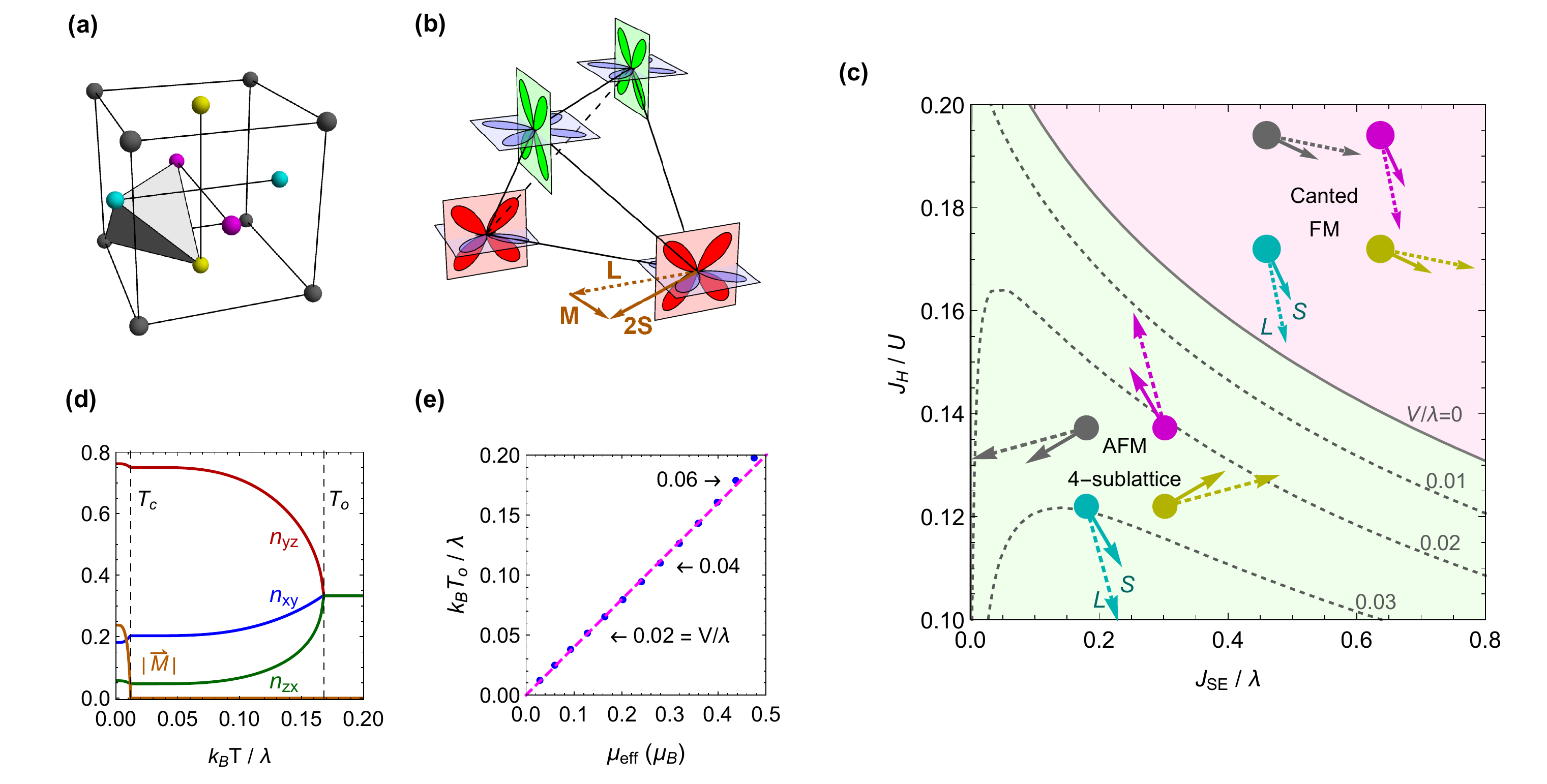}
\caption{
\label{d1_5plots}
(a) FCC lattice decoupled into four inequivalent sites shown by four different colors.
(b) The orbital ordering pattern driven by both $J_\mathrm{SE}$ and $V$ constrains the direction of orbital angular momentum.
Deviations from the $j = 3/2$ limit produce a net magnetic moment $\veck{M}$ as the spin and orbital components separate.
(c) The zero temperature phase diagram shows phases where the moments in each plane of the page (eg. plane containing yellow and black sites) are collinear and the moments between planes are at approximately 90 degrees due to the orbital ordering pattern.
Increasing inter-orbital repulsion $V$ between sites reduces minimum strength of Hund's coupling required to induce FM.
(d) Mean field values for the bottom sites (black, yellow) are shown as a function of temperature.
The $n_{yz}$ orbital (red) has the largest occupancy followed by the $xy$ orbital (blue).
(e) With $J_\mathrm{SE} = 0$, we calculate the orbital ordering temperature $T_o$ and effective Curie moment enhancement $\mu_\mathrm{eff}$ for different values of $V$.
}
\end{figure*}

We decouple $H_\mathrm{SE}$ and $H_\mathrm{V}$ into all possible on-site mean fields, i.e. $\veck{S}_i n_i^\alpha \veck{S}_j n_j^\alpha \rightarrow \veck{S}_i n_i^\alpha \langle \veck{S}_j n_j^\alpha \rangle + \langle \veck{S}_i n_i^\alpha \rangle \veck{S}_j n_j^\alpha - \langle \veck{S}_i n_i^\alpha \rangle \langle\veck{S}_j n_j^\alpha \rangle$.
Since the FCC lattice is not bipartite, we decouple into four inequivalent sites shown in \figref{d1_5plots}(a) where each set of four inequivalent neighbors forms a tetrahedron.
Since the mean fields need not factor into the product of spins and orbitals,  $\langle \veck{S}_i n_i^\alpha\rangle \neq \langle \veck{S}_i \rangle \langle n_i^\alpha \rangle$, there are a total of 15 mean fields per site comprised of three spin operators, three orbital operators, and products of the spin and orbital operators.
Applying the constraint that one electron resides on each site, there are 11 independent mean fields per site giving a total of 44 mean fields in the tetrahedron.
We then numerically solve for the lowest energy solutions of the mean field equations.

\subsection{Zero Temperature Mean Field Theory}

In the limit where spin-orbit coupling $\lambda$ is the dominant energy scale, the magnetically ordered phases can be characterized by an arrangement of ordered $J=3/2$ multipoles.\cite{ChenBalents2010}
However, when $J_\mathrm{SE}$ and $\lambda$ are comparable, a multipolar description within the $j=3/2$ states breaks down and consequently both spin and orbital parts must be considered independently.
Furthermore, the orbital contributions come in the forms of both orbital occupancy $n^\alpha$ and orbital angular momentum $\veck{L}$.
Since $n^\alpha$, $\veck{L}$, and $\veck{S}$ are coupled, there is competition between order parameters which results in non-trivial ordering.

The zero temperature phase diagram is shown in \figref{d1_5plots}(c) as a function of the strength of Hund's coupling $\eta=J_H/U$ and superexchange $J_\mathrm{SE}/\lambda$.
Large values of $\eta$ support a canted ferromagnetic (FM) structure while smaller values support an antiferromagnetic (AFM) structure.
The spin-1/2 and orbital-1 angular momenta order parameters $\langle \veck{S} \rangle$ and $\langle \veck{L} \rangle$ are shown for each of the four inequivalent sites from \figref{d1_5plots}(a).
In both phases, one of the three directions has no ordered angular momenta, e.g. $\langle L_z \rangle = \langle S_z \rangle = 0$, so that both magnetic structures are co-planar.
Both phases feature some separation of the ordered spin and orbital moments which increases as a function of $J_\mathrm{SE} / \lambda$.
To understand why these magnetic structures emerge, we examine the orbital occupancy order parameters, $n^\alpha$, separately from the magnetic order parameters.
In both the FM and AFM phases, there is an orbital ordering pattern pictured in \figref{d1_5plots}(b).
The two sites in the lower plane of \figref{d1_5plots}(b) have the $yz$ orbital (red) with the highest electron occupancy while the $xy$ orbital (blue) receives the second highest and the $zx$ orbital receives the lowest (green, not pictured).
The two sites in the upper plane have identical ordering except the roles of the $yz$ and $zx$ orbitals are reversed.
Qualitatively this orbital ordering pattern is favored by both the $H_\mathrm{V}$ and $H_\mathrm{SE}$ terms which pushes electrons onto orbitals that have small overlaps.
This allows the electron on a green orbital to hop onto an unoccupied green orbital in the plane directly above or below (and similarly for red orbitals).
Since these mechanisms work to suppress the overlap of half filled orbitals, ferromagnetic interactions may become energetically favorable.
A derivation of the mean field solution for $H_\mathrm{V}$ is provided in Appendix \ref{AppendixMomentEnhancement}.

Once orbital order sets in, the allowed magnetic phases are restricted by the direction of orbital angular momentum.
Full orbital polarization is time-reversal invariant and would not allow orbital magnetic order.
However \figref{d1_5plots}(d) shows that each site has at least two orbitals with non-negligible occupancy which allows for the development of an orbital moment.
Thus the direction of the orbital moment is determined by the direction common to the two planes of occupied orbitals with the overall sign of the direction (e.g. $+x$ or $-x$) left undetermined.
Figure \ref{d1_5plots}(c) shows that the orbital angular momenta between planes are close to 90 degrees apart for both FM and AFM phases.
As spin and orbital angular momentum are coupled together, the spin moments will select which direction the orbital moments choose (i.e. $+x$ or $-x$).
The decision to enter an FM or AFM state is then determined by the spin interactions characterized both by the strength of $\eta = J_H / U$ and the magnitude of the orbital order parameter.
If $\eta$ is large, then ferromagnetic spin interactions follow and result in both the spin and orbital degrees of freedom aligning within each $xy$ plane producing a net canted FM structure.
If $\eta$ is small, then antiferromagnetic spin interactions follow which result in the 4-sublattice AFM structure.
We note that the Goodenough-Kanamori-Anderson rules\cite{goodenough1963,Kanamori1959,Anderson1959} are not enough to determine whether FM or AFM is favored, and the interplay between spin-orbit coupling and the anisotropic orbital degrees of freedom play a crucial role in tipping the energy balance one way or the other.

There are two additional factors that determine if the FM or AFM state is selected.
The dominant effect is the degree of orbital polarization.
When the strength of inter-orbital repulsion $V$ is increased, the tendency for orbitals to order becomes stronger.
This disfavors the overlap of half filled orbitals causing AFM superexchange, and hence promotes FM superexchange.
Figure \ref{d1_5plots}(c) shows a dramatic shift toward FM when a small $V$ interaction is included.
The second effect comes from the separation of spin and orbital degrees of freedom.
When $J_\mathrm{SE}$ becomes comparable to $\lambda$, the spin moments can partially break away from the orbital moments tending more toward a regular spin FM state instead of a canted spin FM state.
Since a spin AFM state does not benefit from this separation to the same extent, FM becomes increasingly energetically favorable.

Dimer phases have been proposed \cite{ChenBalents2010,RomhanyiBalentsJackeli2016} and offer a way to explain the absence of magnetic order in $d^1$ materials.
However when $\lambda / J_\mathrm{SE}$ is large, these dimer phases only occur at very small values of $\eta = J_H / U$.\cite{RomhanyiBalentsJackeli2016}
Furthermore, orbital repulsion $V$ acts to further suppress dimerization.
Since our focus is on the magnetically ordered phases of these double perovskites, we will not pursue these possibilities in this work.

\subsection{Finite Temperature Mean Field Theory}

We now examine the model at finite temperature.
Figure \ref{d1_5plots}(d) shows a characteristic order parameter vs temperature curve.
At high temperatures all order parameters are trivial and each orbital occupancy takes a value of $n^{yz} = n^{zx} = n^{xy} = 1/3$.
As temperature is lowered, the first transition is to a time reversal invariant orbitally ordered state [see \figref{d1_5plots}(b)] at temperature $T_o$ whose scale is set both by $V$ and $J_\mathrm{SE}$.
At $T_o$, the entropy released is from orbital degeneracy, even when $V=0$.
Below the second transition at a temperature $T_c$ whose energy scale is set only by $J_\mathrm{SE}$, time reversal symmetry is broken on each site with the development of magnetic order, and the remaining entropy is released.

The fundamental question arises of how large the exchange interaction $J_\mathrm{SE}$ and repulsion $V$ are in materials systems.
Fits to experimental susceptibility\cite{Stitzer2002} show \ce{Ba2LiOsO6} and \ce{Ba2NaOsO6} have relatively small Curie-Weiss temperatures of $\theta = -40.5$ K and $\theta = -32.5$ K respectively indicating that $J_\mathrm{SE}$ in cubic $5d^1$ double perovskites is weak.
However integrated heat capacity\cite{EricksonFisher2007} of \ce{Ba2NaOsO6} shows an entropy release just short of $R \ln 2$ at $T_c$ consistent with the splitting of a local Kramer's doublet with no further anomalies in heat capacity up to 300 K.
%A recent NQR study has also reported evidence for this quadrupolar order at high temperature.CITE XX
This suggests $T_o \gg T_c$ so that $V$ is the most relevant parameter for determining the properties well above $T_c$.

Since the onset of orbital order necessarily alters the angular momenta available to order and respond to an applied magnetic field, we calculate how the effective Curie-Weiss constant depends on orbital ordering.
Using $J_\mathrm{SE} = 0$, we calculate the temperature dependent susceptibility within mean field theory as a function of temperature for different values of $V/\lambda$.
For each value of $V/\lambda$ we calculate both the orbital ordering temperature $T_o$ and the effective Curie moment $\mu_\mathrm{eff} = g \mu_B \sqrt{J(J+1)}$ from a fit to low temperature inverse susceptibility.
\figref{d1_5plots}(e) gives numerical results from our mean field theory that shows a linear relationship between $T_o$ and $\mu_\mathrm{eff}$.
In the absence of orbital order, the projection of the magnetization operator to the $J=3/2$ space is identically zero.
However once orbital order sets in, the $j=1/2$ components of the wavefunction get mixed with the $j=3/2$ components.
The matrix elements that connect these two $J$ spaces then acquire expectation values and allow the effective Curie moment to become non-zero.
An approximate derivation of this relation is provided in Appendix \ref{AppendixMomentEnhancement}.

\begin{figure}
\includegraphics[width=8.5cm,trim=0cm 0cm 0cm 0cm]{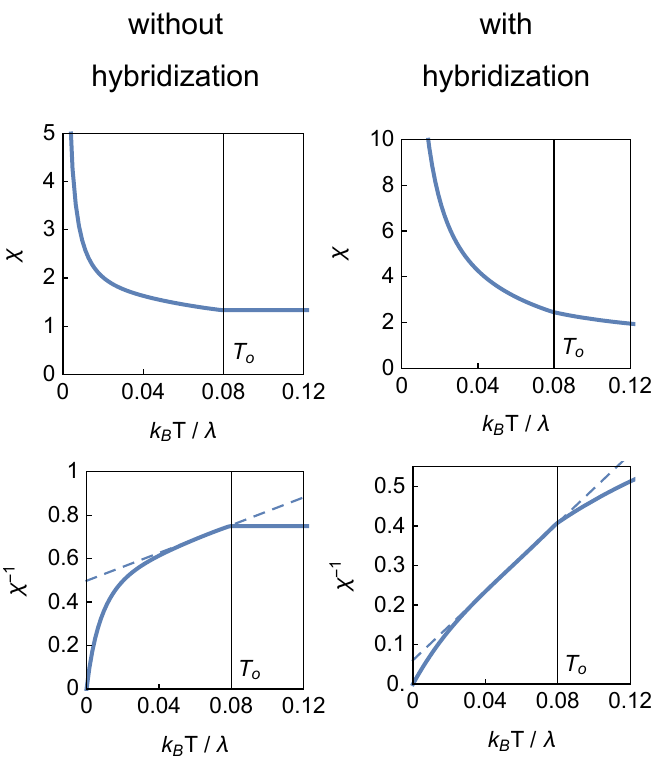}
\caption{
Typical susceptibility, $\chi = \tfrac{1}{3} ( \chi_{xx} + \chi_{yy} + \chi_{zz})$, and inverse susceptibility are plotted against temperature.
The susceptibility curves are shown both without the correction due to hybridization, $\gamma = 1$, and with the correction, $\gamma = 0.536$.
We have chosen $J_\mathrm{SE} = 0$ and left $V$ finite to illustrate the consequence of orbital order on the susceptibility.
By choosing $J_\mathrm{SE} = 0$, we show that although $T_c = 0$ while $T_o \neq 0$, the fitted Curie-Weiss temperature appears to be negative.
Note that a single Curie-Weiss fit cannot span the entire range below $T_o$.
\label{d1_susceptibility}
}
\end{figure}

In addition to the perturbative separation of $L$ and $S$ due to mixing of the $J$ multiplets, hybridization with oxygen has been shown to greatly reduce the orbital contribution to the moment.\cite{AhnKunes2016,XuBrink2016}
Here the magnetization operator assumes the form $\veck{M} = 2\veck{S} - \gamma \veck{L}$ where $\gamma = 0.536$ and results in an effective Curie moment of $0.60 \mu_B$ compared to an experimental value of $0.67\mu_B$.\cite{Stitzer2002}
However the onset of quadrupolar order within the $j=3/2$ states results in a reduction of the nominal $0.60 \mu_B$ value.
In general, the projection of a linear combination of the $n_{yz}$, $n_{zx}$, and $n_{xy}$ operators to the $j=3/2$ states is (up to a constant shift) a linear combination of the operators $J_x^2 - J_y^2$ and $J_z^2$.
By projecting to the lowest energy doublet induced by these operators, we may calculate the $g$ factors for this pseudo-spin 1/2 space.
While the $g$ factors are different in the three cubic directions due to the anisotropic nature of quadrupolar order, the sum of the squares is a constant, and the powder average is $g^2 = \tfrac{1}{3} (g_x^2 + g_y^2 + g_z^2) = 3$.
Then splitting of the $j=3/2$ states reduces the Curie moment by a factor of $(g \sqrt{3/4})/(\sqrt{15/4}) = \sqrt{3/5}$ which makes the calculated moment $0.47\mu_B$.
We find that mixing between the $j=3/2$ and $j=1/2$ states brings the calculated moment closer to experimental values.

There are more consequences of orbital ordering that are particularly important for the magnetic susceptibility of this spin-orbital system.
The orbital order reduces symmetry of the system and causes susceptibility to become anisotropic.
Since the orbital ordering pattern tends to push angular momentum into the ordering planes, susceptibility is enhanced in these two directions while reduced in the third direction.
Although anisotropic susceptibility is expected once cubic symmetry is broken, it is an easy test to determine at what temperature orbital order occurs.
However this is yet a more important effect.
When orbital order sets in at $T_o$, the effective moment changes as the orbital degrees of freedom tend toward a (partially) quenched state which results in an effective moment which changes with temperature.
The non-Curie-Weiss behavior will be critical when interpreting the observed negative Curie-Weiss temperatures in $5d^1$ ferromagnetic compounds.

Within our mean field theory, we now calculate the susceptibility without the hybridization correction $\gamma$ and with the hybridization correction to show this effect.
For clarity, we set $J_\mathrm{SE} = 0$ to isolate the contributions from orbital order from those of magnetic interactions.
Fig.~\ref{d1_susceptibility} shows that below the orbital ordering temperature, the susceptibility deviates from the Curie-Weiss law.
However the data below $T_o$ can be fit over a large range to give a negative Curie-Weiss intercept despite the absence of magnetic interactions.
Although the region where the fit works the best is just below $T_o$ where the orbital occupation is rapidly changing, there is a quantitative explanation for this.

We consider the case without hybridization where the effective moment for the $j=3/2$ states is identically zero.
When orbital order occurs, there is mixing between the $j=3/2$ and $j=1/2$ states proportional to $V\langle \delta n \rangle / \lambda$.
Then below $T_o$, the effective magnetization operator for the lowest energy Kramer's doublet increases in a way proportional to $\langle \delta n \rangle$ due to the matrix elements between $j=3/2$ and $j=1/2$.
The effective Curie moment goes as the square of magnetization and thus the enhancement is of order $\langle \delta n \rangle^2$.
Since orbital order below $T_o$ scales as $\langle \delta n \rangle \propto |T_o-T|^{1/2}$ within mean field theory, the effective Curie moment gains a contribution scaling as $|T_o - T|$ just below $T_o$.
At temperatures far away from $T_c$, the leading correction to susceptibility and and inverse susceptibility is linear leading to the appearance of a Curie-Weiss law.
We note, however, that this is artificial and is not indicative of the physical magnetic interactions.

Despite using mean field critical exponents, qualitatively we have understood how deviations from the Curie-Weiss law occur from changing orbital occupancy.
Because we have used a simple model consisting of only $\lambda$ and $V$ with a-priori knowledge of the ideal Curie-Weiss temperature of zero, we have been able to clearly interpret the non-Curie-Weiss susceptibility.
However the fitting procedure must be performed with some caution since both the fit region and the chosen value of $\chi_0$ (temperature independent term) determine the reported $\theta_\mathrm{CW}$ and $\mu_\mathrm{eff}$.
In fact, experimental behavior may deviate even more strongly due to the quantitative details of how orbital occupancies change with temperature.
In particular, coupling between orbitals and phonons may be a crucial aspect here.\cite{XuBrink2016}

Reference \citenum{ChenBalents2010} claimed negative Curie-Weiss temperatures were achievable in their model for ferromagnetic ground states, although this crucial result was not explicitly shown.
Reference \citenum{Marjerrison2016} has reproduced that model under the circumstances necessary to generate ferromagnets with negative Curie Weiss temperatures, and they find jump discontinuities (finite-to-infinite) in the magnetic susceptibility at $T_c$.
Such jump discontinuities are not seen in \ce{Ba2NaOsO6}, \ce{Ba2MgReO6}, or \ce{Ba2ZnReO6}.
We note that our mechanism for shifting the Curie-Weiss temperature is free from these discontinuities and features a properly diverging susceptibility at $T_c$ for the ferromagnetic phase, thereby providing a more accurate and natural description of the transition.

\section{$d^2$ Double Perovskites}

\begin{figure*}[t]
\includegraphics[width=0.95\textwidth,trim={0.5cm 1cm 0cm 0cm}]{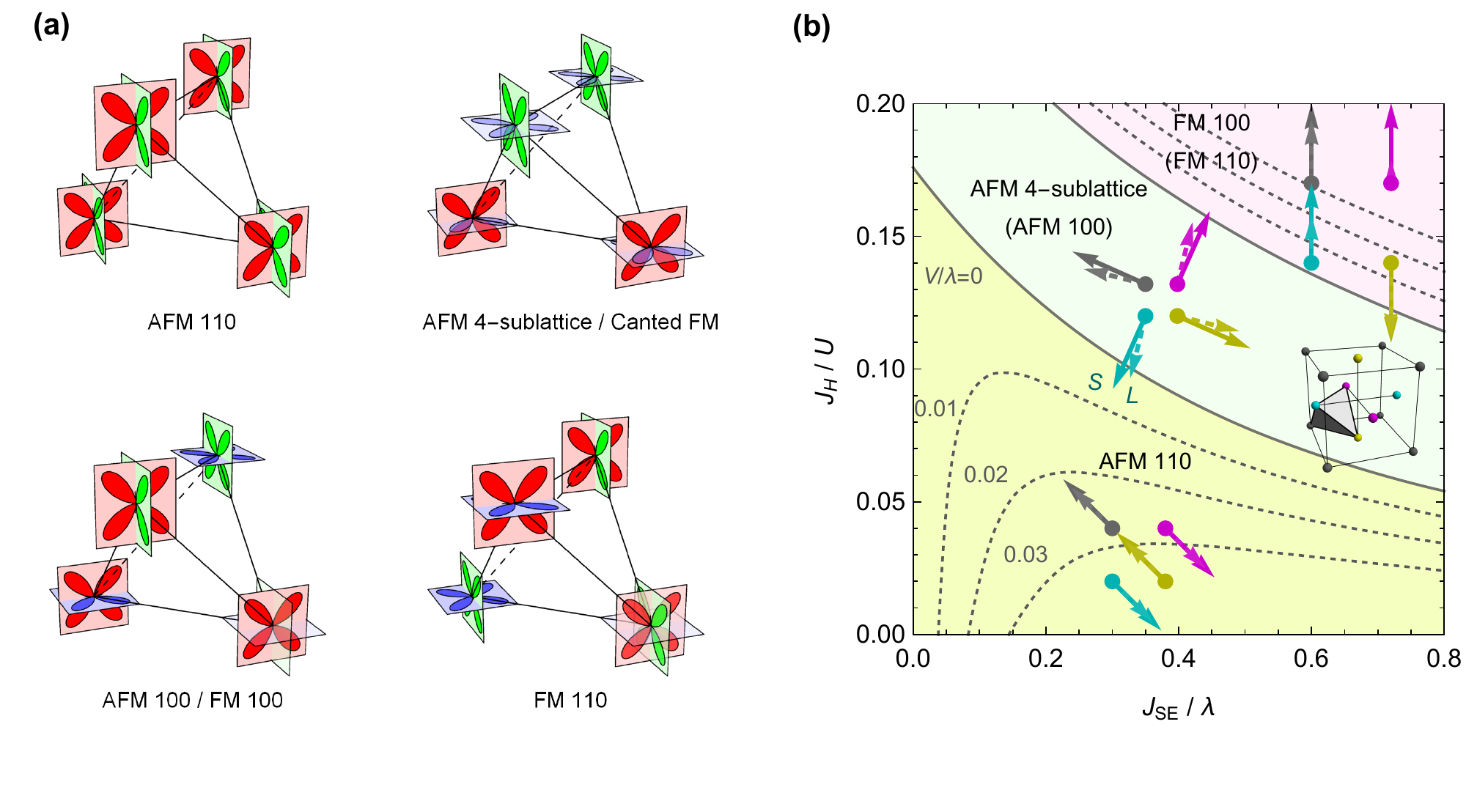}
\caption{
\label{d2_5plots}
(a) Orbital ordering patterns are shown for each type of magnetic order. Orbitals shown in solid colors represent the most occupied orbitals while orbitals not shown or shown transparently have lower occupancy. 
(b) The zero temperature phase diagram shows three ground state phases: AFM with moments (anti)parallel to [110], AFM 4-sublattice structure, and FM with moments parallel to [100].
Phases shown in parenthesis (AFM [100], FM [110]) show the next lowest energy phase in each region.
}
\end{figure*}

Here we will modify the $d^1$ spin-orbital model to accommodate two electrons.
Again, we then solve the model within mean field theory at both zero temperature and finite temperature.
At zero temperature, we find new orbital phases not found in our $d^1$ phase diagram.
For completeness, we show susceptibilities and orbital occupancies at finite temperature.

\subsection{Model}
Our model for $d^2$ is constructed from the same considerations used in $d^1$ only changing the electron count.
The tight-binding model $H_\mathrm{TB}$, the inter-site orbital repulsion $H_\mathrm{V}$, and the on-site Coulomb interaction $H_\mathrm{U}$ are valid for the $d^2$ model without modification.
However spin-orbit coupling and superexchange will change since the total spin and orbital angular momentum on each site are now composed of two electrons.
In the Mott limit, Hund's rules are enforced by $H_\mathrm{U}$ resulting in a total spin $S=1$ and total orbital angular momentum $L=1$ on each lattice site.
Within this space, the spin-orbit interaction takes the form $H_\mathrm{SO}' = -\tfrac{\lambda}{2} \sum_i \veck{L}_i \cdot \veck{S}_i$.
The superexchange Hamiltonian is given by the following
\begin{equation}
\begin{split}
H_\mathrm{SE}' = -\frac{J_\mathrm{SE}}{12} \sum_\alpha \sum_{\langle ij \rangle \in \alpha} \left\lbrace r_1 (2+\veck{S}_i \cdot \veck{S}_j) (n_i^\alpha - n_j^\alpha)^2 \right. \\
\left.  (1 - \veck{S}_i \cdot \veck{S}_j) \left[ (n_i^\alpha + n_j^\alpha)^2 + (\tfrac{3}{2} r_3 - \tfrac{5}{2})n_i^\alpha n_j^\alpha \right] \right\rbrace
\end{split} 
\label{HSEd2}
\end{equation}
where the definitions of $J_\mathrm{SE}$, $r_1$, and $r_3$ correspond to those used previously.
As before, the top line in \eqref{HSEd2} gives a ferromagnetic spin interaction when only one of the two interacting orbitals is occupied while the second line gives an antiferromagnetic spin interaction which is maximized when two half filled orbitals overlap.
The total effective magnetic interaction then reads $H' =H_\mathrm{SO}' + H_\mathrm{SE}' + H_\mathrm{V} $.
We decouple $H_\mathrm{SE}'$ and $H_\mathrm{V}$ into all possible on-site mean fields using four inequivalent sites as before and then solve the mean field equations numerically.

\subsection{Zero Temperature Mean Field Theory}

The zero temperature phase diagram is shown in \figref{d2_5plots}(b) as a function of the strength of Hund's coupling $\eta = J_H / U$ and superexchange $J_\mathrm{SE}/\lambda$.
In the limit of large spin-orbit coupling and the absence of inter-site orbital repulsion, the ground state is predominantly AFM with the moment aligning parallel to the [110] direction within a plane and antiparallel to the [110] direction in the next plane.
To see why this phase occupies such a large region of phase space, we analyze the orbital structure that accompanies it, as shown in \figref{d2_5plots}(a).
On each site, one electron moves onto the $yz$ orbital and the other onto the $zx$ orbital.
In this configuration both occupied orbitals overlap with occupied orbitals on neighboring sites and unoccupied orbitals overlap with other unoccupied orbitals so that AFM superexchange is maximized.
These orbitally controlled AFM interactions then take place between planes and not within planes resulting in AFM between planes while FM interactions prevail in each plane.
Since this this orbital pattern is compatible with tetragonal distortion, as observed in \ce{Sr2MgOsO6}\cite{Morrow2016}, we expect nominally cubic crystal structures to distort.

The next phase we find is the AFM 4-sublattice coplanar structure previously found in the $d^1$ phase diagram.
As before, the orbital degrees of freedom are closely aligned with the directions perpendicular to the occupied orbitals, and the spin and orbital moments perturbatively separate from each other with increasing superexchange.
It is worth noting that in this region of the phase diagram, the next lowest energy phase is AFM [100] that can become a competitive ground state upon inclusion of anisotropy.

For large superexchange and Hund's coupling, we find a ferromagnetic phase with ordering along the [100] direction that is best characterized as a ``3-up, 1-down'' collinear structure where three of the four moments order parallel to each other along the chosen direction and the fourth moment orders anti-parallel to the other three.
It is worth noting that the second most energetically favorable phase in this region of the phase diagram is another ``3-up, 1-down'' structure where each moment is either approximately parallel or antiparallel to the [110] direction.
The energy difference between the FM [100] and FM [110] phases is negligible and either phase is a suitable ground state.
In addition to these two FM phases, we find a canted FM solution to the mean field equations with the same orbital ordering pattern as the $d^1$ canted FM phase.
However it is higher enough in energy to rule out as a viable ground state and consequently is not shown in the phase diagram.

Unlike the AFM [110] and AFM 4-sublattice structures, the FM/AFM [100] structures features an approximately higher degree of degeneracy due to the orbital degrees of freedom.
Like the AFM 4-sublattice orbital structure, the FM/AFM [100] orbital structure tends to minimize repulsion between orbitals.
Of the four tetrahedral sites, three of them are able to minimize the repulsion and allow occupied orbitals to hop to unoccupied orbitals.
While the repulsion is minimized between those three sites, this forces occupied orbitals on each site to point toward the fourth site.
Figure \ref{d2_5plots}(a) shows that this fourth site in the FM/AFM [100] orbital pattern chooses one of the orbitals to have a majority occupancy (solid color) and the other two orbitals to have minority occupancies (semi-transparent colors).
In the FM [110] phase, a similar situation occurs with the main difference being that now two orbitals have majority occupancy and one orbital has minority occupancy.
Before magnetic order sets in, the degeneracy is approximately extensive as the fourth site on every tetrahedron in the lattice has local orbital frustration.

When inter-site orbital repulsion $H_\mathrm{V}$ is included, the phase boundaries shift.
The most dramatic effect is the recession of the boundary between AFM [110] and the AFM 4-sublattice structure.
This becomes apparent by comparing the orbital configurations of the two phases as the AFM [110] structure maximizes the number of AFM singlets which are penalized by the orbital repulsion.
Unlike in the $d^1$ situation, we find that the inclusion of $V$ does not enhance FM.
While the FM/AFM [100] and FM [110] orbital structures are much more compatible with $H_\mathrm{V}$ than the AFM [110] structure, the AFM 4-sublattice structure still dominates.
We note that unlike the $d^1$ case, canted FM is not favorable here due to the electron count.
The $d^1$ case relies on pushing the large majority of the electron weight onto one orbital while retaining a smaller occupancy on a second orbital to generate an orbital moment.
However in $d^2$, this second orbital must also be occupied which consequently induces AFM interactions within each horizontal plane.

Although we have focused on spin-orbital magnetic order, it is necessary to remark that exotic singlet ground states are also possible.
The Kramer's theorem guarantees that trivial ionic singlets will not occur in $d^1$ systems, and therefore the experimental observation of singlet behavior is an indication of a non-trivial ground state.
Such considerations do not apply to $d^2$, and experimental observations of singlet behavior may arise from trivial local magnetic singlets.
Consequently this local non-magnetic singlet possibility must first be ruled out when searching for exotic singlet behavior.

\subsection{Finite Temperature Mean Field Theory}

\begin{figure*}
\includegraphics[width=\textwidth,trim={0cm 0cm 0.3cm 0cm}]{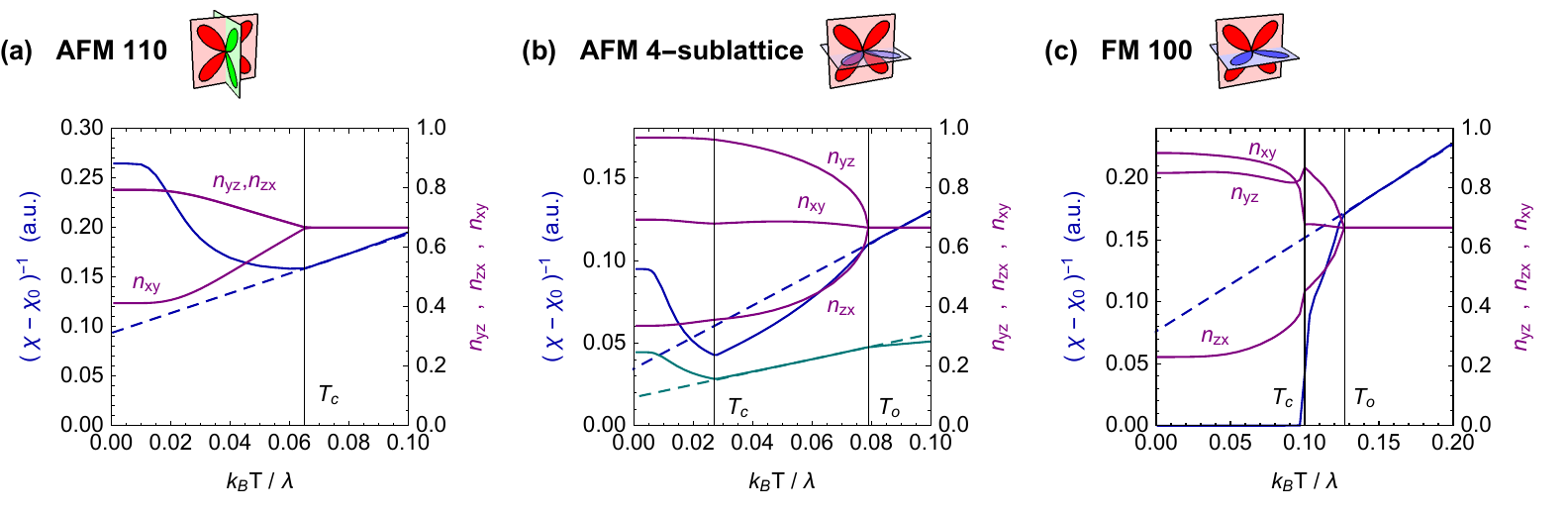}
\caption{
\label{d2Suscept}
Characteristic inverse susceptibility (blue) and orbital occupation (purple) curves are plotted against temperature for the three phases in \figref{d2_5plots}: (a) AFM [110], (b) AFM 4-sublattice, and (c) FM [100].
Susceptibility is averaged over all three directions, $\chi^{-1} = 3 (\chi_{xx} + \chi_{yy} + \chi_{zz})^{-1}$, and all sites in the tetrahedra.
Orbital occupancies are shown for the site pictured above each plot.
}
\end{figure*}

Here we consider the model at finite temperature.
Figure \ref{d2Suscept} shows orbital occupations and inverse magnetic susceptibility as a function of temperature for the three ground state phases from the previous section.
At high temperature, the orbitals have a uniform occupancy of $n^{yz} = n^{zx} = n^{xy} = 2/3$.
There is a temperature $T_o$ where time-reversal invariant order sets in through the orbitals and second temperature $T_c$ where magnetic order sets in.
In the case of the AFM [110] phase, \figref{d2Suscept}(a) shows the two ordering temperatures coincide and that the electrons are pushed onto the $n_{yz}$ and $n_{zx}$ orbitals to maximize antiferromagnetic superexchange.
This is different from the orbital ordering previously reported because this ordering maximizes orbital repulsion instead of minimizing it, so orbital order itself is not favorable and is entirely driven by antiferromagnetic superexchange.
In this situation, the Curie-Weiss law with a negative Curie-Weiss temperature occurs as expected.

The transition to an AFM 4-sublattice structure is shown in \figref{d2Suscept}{(b)}.
Above $T_o$ susceptibility follows the Curie-Weiss law with a negative Curie-Weiss constant.
Below $T_o$ the orbital occupancies change along with the inverse susceptibility to deviate from the high temperature behavior.
Just below $T_o$, susceptibility may be fit to another Curie-Weiss law with another negative Curie-Weiss constant.
Similarly to the $d^1$ case, there is still deviation from the Curie-Weiss law in this regime, however, the deviations are smaller and so is the enhancement of the effective magnetic moment due to mixing of the $J=2$ states with higher energy multiplets.
But we note that when $J_\mathrm{SE} = 0$, we still find the appearance of a negative Curie-Weiss constant due to non-Curie-Weiss susceptibility as we did in the $d^1$ model.

Finally, the transition to an FM [100] structure is shown in \figref{d2Suscept}{(c)}.
Deviations from the Curie-Weiss law are seen below $T_o$, and the sign of the Curie-Weiss constant can switch from negative to positive depending which region fitted.
Unlike the other phases, magnetic order appears at $T_c$ with a first-order transition marked by the jumps in orbital occupancy and susceptibility.
This arises from competition between having the most energetically favorable orbital structure at high temperature and the most energetically favorable magnetic structure at low temperature.

As in the $d^1$ case, we compare values of the theoretical moments to those from experiment.
Oxygen hybridization will result in a Curie moment of $\mu_\mathrm{eff} = \sqrt{6}(1-\gamma / 2)\mu_B$.
Assuming almost half of the moment resides on oxygen, the calculated moment is then $\mu_\mathrm{eff} \approx 1.8 \mu_B$.
This is close to the experimentally observed moments in \ce{Sr2MgOsO6} and \ce{Ca2MgOsO6} (both $1.87\mu_B$)\cite{Yuan2015} but further off from those of \ce{Ba2YReO6} ($1.93\mu_B$)\cite{Aharen2010d2} and \ce{La2LiReO6} ($1.97\mu_B$)\cite{Aharen2010d2}.

\section{Conclusions}
We have studied spin-orbital models for both $d^1$ and $d^2$ double perovskites where the B' ions are magnetic and have strong spin-orbit coupling.
We found several non-trivial magnetically ordered phases characterized both by ordering of the spin/orbital angular momentum and ordering of the orbitals.
This orbital ordering shows why ferromagnetism is energetically favorable in these systems when electron count is $d^1$ but not when it is $d^2$, particularly at large spin-orbit coupling.
Additionally, ordering of the orbital degrees of freedom can produce non-Curie-Weiss behavior which can lead to the appearance of a negative Curie-Weiss in the canted ferromagnetic phase.
We emphasize that examination of the spin and orbital degrees of freedom separately gives an enhanced qualitative understanding of the magnetism for this class of spin-orbit coupled double perovskites.

\section{Acknowledgements}
We thank Patrick Woodward and Jie Xiong for their useful discussions.
We acknowledge the support of the Center for Emergent Materials, an NSF MRSEC, under Award Number DMR-1420451.

\appendix
\section{$\mu_\mathrm{eff}$ enhancement and $T_o$ for $d^1$ model}
\label{AppendixMomentEnhancement}
To obtain the orbital ordering temperature $T_o$ and the effective moment $\mu_\mathrm{eff}$ as a function of $V/\lambda$, we will solve the mean field equations for $H_\mathrm{V} + H_\mathrm{SO}$ analytically.
The relevant mean field parameters for the four sites from \figref{d1_5plots}(b) are given below
\begin{equation}
\langle n^{xy}_{ 1} \rangle =
\langle n^{xy}_{ 2} \rangle =
\langle n^{xy}_{ 3} \rangle =
\langle n^{xy}_{ 4} \rangle = \tfrac{1}{3} + \delta n_z
\end{equation}
\begin{equation}
\langle n^{yz}_{ 1} \rangle =
\langle n^{yz}_{ 2} \rangle  =
\langle n^{zx}_{ 3} \rangle =
\langle n^{zx}_{ 4} \rangle = \tfrac{1}{3} + \delta n_x
\end{equation}
with the condition $\sum_\alpha n_i^\alpha = 1$ determining the other four parameters.
We obtain the single site mean field Hamiltonian for $V$.
\begin{equation}
H_\mathrm{V}' = -V \left[ ( \tfrac{86}{3}  \delta n_x + \tfrac{43}{3} \delta n_z ) n^{yz} + ( \tfrac{43}{3}  \delta n_x + \tfrac{53}{3} \delta n_z ) n^{xy} \right]
\end{equation}
Since above $T_c$, the high mean field Hamiltonian $H_\mathrm{MF}' = H_\mathrm{V}' + H_\mathrm{SO}$ is time reversal invariant, we rotate into the basis of total angular momentum $J$ which factors into two $3 \times 3$ blocks of doublets.
The upper block may be chosen to be of the form below
\begin{equation}
 \left(
\begin{array}{ccc}
 \tfrac{3\lambda}{2} & -\frac{43 V (2 \delta n_x +\delta n_z )}{3 \sqrt{6}} & -\frac{7 V \delta n_z }{\sqrt{2}} \\
 -\frac{43 V (2 \delta n_x +\delta n_z )}{3 \sqrt{6}} & \frac{7 V \delta n_z }{2}
   & \frac{43 V(2 \delta n_x + \delta n_z )}{6 \sqrt{3}} \\
 -\frac{7 V \delta n_z }{\sqrt{2}} & \frac{43 V (2 \delta n_x + \delta n_z )}{6
   \sqrt{3}} & -\frac{7 V \delta n_z }{2} \\
\end{array}
\right)
\end{equation}
where the basis $\ket{J,m_J}$ is given by $\ket{1/2,+1/2}$, $\ket{3/2,-3/2}$, $\ket{3/2,+1/2}$ in this order.
Using $\theta = \arctan 43 \sqrt{3} \left(2 \delta n_x + \delta n_z \right) / 63 \delta n_z$, we diagonalize the Hamiltonian in the $j=3/2$ block
\begin{equation}
\left( \begin{array}{ccc}
\tfrac{3\lambda}{2} & x & y \\ 
x & -\Delta & 0 \\ 
y & 0 & \Delta 
\end{array} \right)
\end{equation}
where 
\begin{equation}
\Delta = V \frac{ \sqrt{1849 \delta n_x (\delta n_x + \delta n_z )+793 \delta n_z^2} }{ 3 \sqrt{3} }
\end{equation}
\begin{equation}
x = -V \frac{ 43 \sqrt{3} (2 \delta n_x + \delta n_z ) \cos \tfrac{\theta}{2} +63 \delta n_z \sin \tfrac{\theta}{2} }{9 \sqrt{2}}
\end{equation}
and $y$ is given by $x$ with $\sin \theta \rightarrow \cos \theta$ and $\cos \theta \rightarrow -\sin \theta$ applied.
The lowest $J=3/2$ doublet with energy $-\Delta$ is mixed with the $J=1/2$ doublet with amplitude $-2x/3\lambda$.

We project the magnetization operator $\veck{M} = 2\veck{S} - \veck{L}$ onto this lowest doublet.
Since nominally $g=0$ for the $j=3/2$ states, the first non-zero correction to the wavefunction comes from mixing of the $j=3/2$ and $j=1/2$ states.
From the projection, we obtain the $g$ factors for this doublet in all three directions (ie. $M_x = g_x \tfrac{\mu_B}{2} \sigma_x$, etc) and compute the average $g$ factor obtained in a powder susceptibility measurement $g^2 = \frac{1}{3} \left( g_x^2 + g_y^2 + g_z^2 \right)$ to obtain the powder average effective moment for the doublet.
For the parameter regime we are interested in, $\delta n_z$ has a negligible contribution to $g$, and the $g$ factor is given approximately by $g=344 V |\delta n_x| / 9 \sqrt{3} \lambda$ so that the moment is $\mu_\mathrm{eff} = 172  V |\delta n_x| \mu_B / 9 \lambda $.

Now we obtain the mean field orbital ordering temperature $T_o$ which occurs when the $j=3/2$ states split.
In the limit that $\delta n_z$ is negligible, we self consistently solve for the expectation value of the operator the projections of the operator $\delta n_x \rightarrow n^{yz} - \tfrac{1}{3}$ within the $2\times 2$ subspace of energies $-\Delta$ and $\Delta$ (ie. $\ket{J=3/2,J_z=-3/2}$ and $\ket{J=3/2,J_z=+1/2}$).
The projection of the $\delta n_x$ operator to this subspace is
\begin{equation}
\delta n_x \rightarrow \left( \begin{array}{ccc}
 -\tfrac{1}{2\sqrt{3}} & -\tfrac{1}{6} \\
 -\tfrac{1}{6} & \tfrac{1}{2\sqrt{3}}
\end{array} \right)
\end{equation}
so that the mean field equations for $\delta n_x$ read
\begin{equation}
\delta n_x = \frac{1}{2\sqrt{3}} \tanh \beta \Delta
\end{equation}
where $\Delta \approx \tfrac{43 V}{3\sqrt{3}} \delta n_x $.
Then we find $k_B T_o = 43 V / 18$ which is consistent with Ref.~ \citenum{ChenBalents2010}.
However, in contrast to Ref.~\citenum{ChenBalents2010}, our analysis shows that this orbital order is compatible with \textit{both} the FM and AFM phases and does \textit{not} disappear below $T_c$ for the AFM phase.
We can relate the ratios of these results as seen in \figref{d1_5plots}(e) by
\begin{equation}
\frac{k_B T_o / \lambda}{\mu_\mathrm{eff} / \mu_B} = \frac{1}{8 \,\delta n_x}.
\end{equation}
Using the zeroth order approximation for $\delta n_x$ as $1/2\sqrt{3}$, this ratio becomes 0.43 which is close to that shown in \figref{d1_5plots}(e).

\newpage 
\bibliography{include}

\end{document}